\documentclass[12pt,onecolumn,draft]{IEEEtran} 

\usepackage{amsmath,amssymb,amsfonts}
\usepackage[final]{graphicx}

\newtheorem{lemma}{Lemma}

\def \M{\mathcal{M}}
\def \G{\mathcal{G}}

\def \L {{\mathcal L }}

\def \E{{\textrm{E}}}

\def\cov{\textrm{cov}}
\def\var{\textrm{var}}

\begin{document}
\title{\Large A Note on the Sum of Correlated Gamma Random Variables
\thanks{*This work is partially supported by the Spanish Government under project TEC2007-67289/TCM
and by AT4 wireless.}}

\author{
\vspace{0mm}
\authorblockN{ Jos\'e F. Paris}}


\maketitle
\begin{abstract}
The sum of correlated gamma random variables appears in the
analysis of many wireless communications systems, e.g. in systems
under Nakagami-$m$ fading. In this Letter we obtain exact
expressions for the probability density function (PDF) and the
cumulative distribution function (CDF) of the sum of arbitrarily
correlated gamma variables in terms of certain Lauricella
functions.
\end{abstract}

\vspace{0mm}

\IEEEpeerreviewmaketitle

\section{Introduction}

\PARstart{M}{any} of the performance analysis problems in the
scope of wireless communications theory require determination of
the statistics of the sum of the squared envelopes of Nakagami-$m$
faded signals or, equivalently, the sum of gamma random variables
since the square of a Nakagami-$m$ random variable follows a gamma
distribution \cite{Alouini2001}.

Some expressions are available in literature for the probability
density function (PDF) of the sum of gamma random variables, e.g.
see \cite{Alouini2001} and the references cited herein. These
expressions are frequently in the form of infinite series, including the
general expression for arbitrary correlation provided in
\cite{Alouini2001}. In this Letter we revisit the result derived
in \cite{Alouini2001}; we show that, under the same assumptions,
the PDF and the cumulative distribution function (CDF) of the sum of correlated gamma random
variables can be expressed in a compact form by certain Lauricella
functions. To the best of the author's knowledge, the
expressions obtained here are novel.

\section{Analytical Results}
\label{resultados}

For clarity we will use the notation adopted in
\cite{Alouini2001}. We say $X$ follows a gamma distribution with
parameters $\alpha>0$ and $\beta>0$ if the PDF of $X$ is given by
\begin{equation}
\label{PDF_gamma}
p_X(x)=\frac{x^{\alpha-1}e^{-x/\beta}}
{\beta^{\alpha}\Gamma(\alpha)}U(x),
\end{equation}
where $\Gamma(\cdot)$ is the gamma function and $U(\cdot)$ is
the unit step function. The shorthand
notation $X\sim \G (\alpha,\beta)$ will be used to denote that $X$ is gamma distributed with parameters
$\alpha$ \mbox{and $\beta$.}

The key idea of this Letter is the following. In
\cite{Alouini2001} the authors extended the Moschopoulos' Theorem \cite{Moscho1985}.
Interestingly, they observed the similarity between the MGF of the correlated case
and the independent case. Then, the Moschopoulos technique of
inverting the MGF was again adopted for the correlated case. Here,
we start from the MGF of the correlated case, but instead of using
the Moschopoulos technique, we establish a connection between the
MGF and certain Lauricella functions.  The first who established a connection of this type was Kabe
\cite{Kabe1962}, within the context of independent gamma random variables. The mathematically precise
statements are given below.

\begin{lemma}
\label{lemma1}
Let $\{X_n\}_{n=1}^{N}$ be a set of $N$ correlated not necessary identically distributed gamma random variables with parameters
$\alpha$ and $\beta_n$, respectively, [i.e., $X_n\sim \G(\alpha,\beta_n)$] and let $\rho_{ij}$ denote the correlation coefficient between $X_i$
and $X_j$, i.e.,
\begin{equation}
\begin{gathered}
  \rho_{ij}  = \rho_{ji}  = \frac{{\cov\left( {X_i ,X_j } \right)}}
{{\sqrt {\var\left( {X_i } \right)\var\left( {X_j } \right)} }},\quad 0 \leqslant \rho_{ij}  \leqslant 1 \hfill \\
  \quad\quad\quad i,j = 1,2, \ldots ,N \hfill \\
\end{gathered}
\end{equation}
then the CDF of $Y = \sum\nolimits_{n = 1}^N {X_n }$ can be
expressed as
\begin{equation}
\begin{gathered}
  F_Y  \left( y \right) = \frac{{y^{N\alpha } }}
{{\det (A)^\alpha  \Gamma \left( {1 + N\alpha } \right)}} \hfill \\
  \quad  \times \Phi _2^{(N)} \left( {\alpha , \ldots ,\alpha ;1 + N\alpha ; - \frac{y}
{{\lambda _1 }}, \ldots , - \frac{y}
{{\lambda _n }}} \right), \hfill \\
\end{gathered}
\end{equation}
where $\Phi _2^{(N)}$ is the confluent Lauricella function \cite{Exton1976}\cite{Erdelyi1954}, and
$\{{\lambda}_n\}_{n=1}^{N}$ are the eigenvalues of the matrix
$A=DC$ where $D$ is the $N\times N$ diagonal matrix with the
entries $\{{\beta}_n\}_{n=1}^{N}$ and $C$ is the $N\times N$
positive definite matrix defined by
\begin{equation}
C = \left( {\begin{array}{*{20}c}
   1 & {\sqrt {\rho _{12} } } &  \cdots  & {\sqrt {\rho _{1N} } }  \\
   {\sqrt {\rho _{21} } } & 1 &  \cdots  & {\sqrt {\rho _{2N} } }  \\
    \vdots  &  \vdots  &  \ddots  &  \vdots   \\
   {\sqrt {\rho _{N1} } } &  \cdots  &  \cdots  & 1  \\

 \end{array} } \right).
\end{equation}
The PDF of $Y$ is given by
\begin{equation}
\begin{gathered}
  f_Y \left( y \right) = \left\{ {\frac{{y^{ - 1 + N\alpha } }}
{{\det (A)^\alpha  \Gamma \left( {N\alpha } \right)}}} \right\} \hfill \\
  \quad  \times \Phi _2^{(N)} \left( {\alpha , \ldots ,\alpha ;N\alpha ; - \frac{y}
{{\lambda _1 }}, \ldots , - \frac{y}
{{\lambda _n }}} \right). \hfill \\
\end{gathered}
\end{equation}
\end{lemma}
\vspace{5mm}
\begin{proof}
See Appendix I.
\end{proof}
\vspace{5mm}
The expressions derived in Lemma \ref{lemma1} are compact and can be
frequently reduced to simpler forms using the properties of the
function $\Phi _2^{(N)}$. In particular, since $\Phi
_2^{(1)}\equiv {}_\text{1}F_1 $, [i.e., equivalent to the
confluent hypergeometric function]  one can check that for $
\alpha  = m$, $\beta _1  = \bar \gamma /m$ and $N=1$ the
expressions derived here reduce to the well-known CDF and PDF of
the square of a Nakagami-$m$ random variable. For reduction formulas, integral representations
and integrals involving $\Phi _2^{(N)}$, the reader should refer to \cite{Exton1976}. Note that
the CDF expression given \mbox{Lemma \ref{lemma1}} allows us to compute the outage probability of
maximal ratio combining (MRC) over correlated Nakagami-$m$ fading channels.

\section{Conclusions}
\label{conclusiones}

In this Letter, compact expressions have been derived for the sum
of arbitrarily correlated gamma random variables. Such expressions
have both theoretical and practical value, and are applicable in a
vast range of wireless communications problems.

\appendices

\section{Proof of Lemma 1}

For an arbitrary function $\phi(x)$ we denote the Laplace transform as $\L[\phi(x);s]$.
As in \cite{Alouini2001}, we define the MGF of $Y$ as
$\M_{Y}(s)=\E[e^{s y}]=\L[f_{Y}(y);-s]$. We know that the MGF of $Y$ is given by  \cite{Alouini2001}
\begin{equation}
M_{Y}(s)=\prod\limits_{n = 1}^N {\left( {1 - \lambda _n s} \right)} ^{ - \alpha }.
\end{equation}
Therefore, we can write the CDF of $Y$ in the following form
\begin{equation}
\label{eq1}
\begin{gathered}
  \L[F_Y \left( y \right);s] = \frac{1}
{s}\L[f_Y \left( y \right);s] \hfill \\
   = \frac{1}
{s}\prod\limits_{n = 1}^N {\left( {1 + \lambda _n s} \right)} ^{ - \alpha }  \hfill \\
   = \left\{ {\frac{{\prod\limits_{n = 1}^N {\left( {\frac{1}
{{\lambda _n }}} \right)^\alpha  } }}
{{\Gamma \left( {1 + \sum\limits_{n = 1}^N \alpha  } \right)}}} \right\}\left\{ {\frac{{\Gamma \left( {1 + \sum\limits_{i = 1}^N \alpha  } \right)}}
{{s^{1 + \sum\limits_{n = 1}^N \alpha  } }}} \right\} \hfill \\
  \quad  \times \prod\limits_{n = 1}^N {\left( {1 - \frac{{( - \frac{1}
{{\lambda _n }})}}
{s}} \right)} ^{ - \alpha }.  \hfill \\
\end{gathered}
\end{equation}
Then, after identifying (\ref{eq1}) with \mbox{\cite[p. 222, eq. 5]{Erdelyi1954},} the CDF of $Y$ is obtained.
To derive the expression for the PDF we can write
\begin{equation}
\label{eq2}
\begin{gathered}
\begin{gathered}
  \L[f_Y \left( y \right);s] = \prod\limits_{n = 1}^N {\left( {1 + \lambda _n s} \right)} ^{ - \alpha }  \hfill \\
   = \left\{ {\frac{{\prod\limits_{n = 1}^N {\left( {\frac{1}
{{\lambda _n }}} \right)^\alpha  } }}
{{\Gamma \left( {\sum\limits_{n = 1}^N \alpha  } \right)}}} \right\}\left\{ {\frac{{\Gamma \left( {\sum\limits_{i = 1}^N \alpha  } \right)}}
{{s^{\sum\limits_{n = 1}^N \alpha  } }}} \right\} \hfill \\
  \quad  \times \prod\limits_{n = 1}^N {\left( {1 - \frac{{( - \frac{1}
{{\lambda _n }})}}
{s}} \right)} ^{ - \alpha }.  \hfill \\
\end{gathered}
\end{gathered}
\end{equation}
Again, after identifying this expression with \mbox{\cite[p. 222, eq. 5]{Erdelyi1954},} the PDF of $Y$ is obtained.


\end{document}